\documentstyle[psfig]{l-aa}

\newcommand{\Msolar}{\mbox{\,$\rm M_{\odot}$}}        
\begin{document}
\thesaurus{1(08.05.1; 08.19.2; 08.16.3; 10.07.3 M~15)}
\title{Distance, structure and bright stellar content
 of the dwarf irregular galaxy UGC 685}

\author{Ulrich Hopp
\inst{1}
\thanks{Visiting Astronomer, German-Spanish Astronomical Center, Calar
Alto, operated by the Max-Planck-Institut f\"ur Astronomie jointly
with the Spanish National Commission for Astronomy}
}
\offprints{U. Hopp}
\institute{Universit\"atssternwarte M\"unchen, Scheiner Str. 1, D 81679 
M\"unchen, FRG, email: hopp~.at.~usm.uni-muenchen.de
}
\maketitle
\begin{abstract}

B and R CCD images and J NICMOS3 frames taken with the Calar Alto 3.5m
telescope of the dwarf irregular galaxy UGC 685 are presented. The
brightest part of the stellar population is resolved in B and R, very few
also in J. The stellar color-magnitude diagram is discussed.  An
estimate of the distance to UGC 685 of $5.5$ Mpc is derived
based on the brightest blue supergiant stars. Most of the resolved
bright stars show colors in the range $-0.1 \le B-R \le +0.7$.

The continuum light of the unresolved stars is distributed rather
regular in B, R, and J, showing only minor irregularities.  This
surface brightness distribution follows an exponential law in all
three colors with a scale length $r_c = 0.33$ kpc. The central surface
brightnesses are $21.57 \pm 0.09 mag/\Box''$, $20.65 \pm
0.06mag/\Box''$, and $20.11 \pm 0.11 mag/\Box''$, in B, R, and J, 
respectively.  The surface
brightness can be traced out to 5 $r_c$ in B and R. Thus, UGC 685
belongs to the class of dwarfs where the HI distribution is much more
extended (here 2.6 times) than the optical (stellar) light
distribution, but contrary to many objects of this type, it does not
belong to the class of low-surface brightness objects. No color
gradient was detected in UGC 685 except that the very center is
slightly bluer. The overall colors are $B-R = + 0.97$, $B-J = +1.55$
and the magnitude is B = 14.55 ($M_B^o$ = -14.5).

The classification as an irregular dwarf from survey plates results from
the few HII regions of UGC 685 which are all concentrated to the
South-East of the center of the galaxy.  On an H${\alpha}$ Calar Alto
2.2m telescope CCD image, I identified only 5 HII regions, one of them
being rather bright. The total H${\alpha}$ flux corresponds to a mildly
on-going star formation with a rate of 0.003 \Msolar yr$^{-1}$, a
low rate even in comparison to other dwarfs. 
The available (and limited) data do not indicate any major deviation
from this rather low star formation rate within the last $10^9$ yr.

\end{abstract}

\thesaurus{1(fill in)}

\section{Introduction}

Isolated dwarf irregular galaxies are good laboratories to study the
star formation history as well as the evolution of galaxies in the
absence of external triggers and without the influence of large-scale
internal triggers like density waves (see Hunter, 1997, for a review).
Understanding local dwarfs constrains the models which were developed
for a description of the large population of sub-luminous, irregular
systems at medium redshifts (see Babul \& Ferguson, 1996; Ferguson \&
Babul 1998). Very nearby galaxies of this type are best suited for
detailed analysis of their star formation histories as they can be
resolved into individual stars (see e.g. Gallart et al, 1996, their table 1,
and Greggio, 1994 and references in these papers). Inside the Local
Group (distance up to roughly 1 Mpc), ground-based deep images under
excellent seeing conditions can be used for this type of investigation
while for more distant galaxies (up to about roughly 5-10 Mpc), HST
data are essential (e.g. Schulte-Ladbeck et al., 1998). For galaxies
in this distance range, radial velocities are very crude distance
estimators as the peculiar velocities are in the order of or larger than
the Hubble flow. A quick and in-expensive method to reject more
distant interlopers from the 10 Mpc sample is the identification of the
brightest stars which can be done with ground-based images of
good seeing (see the discussion by Karachentsev \& Tikhonov, 1993,
K\&T hereafter), and Rozanski \& Rowen-Robinson, 1994, R\&RR). Further,
the ground based images indicate the regions of recent high star
formation activities (HII regions, star clusters), thus allowing a
pre-selection for the pointing of the HST or adaptive optic
observations with their rather small field of view.

UGC 685 (=CGCG0104.7+1625), which belongs to the 10 Mpc sample of 
Kran-Korteweg \& Tammann (1979), was classified as a late type dwarf galaxy
(Sm).  Kran-Korteweg \& Tammann could not associate UGC 685 to one of
their groups and therefore called it a field galaxy. This
indicates that at least the recent evolution and star formation
history of UGC 685 was free of important triggers by interaction,
albeit small mass HI companions as found in several cases by Taylor et
al. (1996) can not be ruled out. The available optical
data as recently compiled by Schmidt \& Boller (1992a,b) indicate a
rather normal and absolutely faint dwarf galaxy at a systemic velocity
(corrected for the local group flow) of $377 km s^{-1}$. The galaxy
was included in various HI surveys which indicate an amount of 
$8.5\cdot10^7$~\Msolar\ in neutral gas and a ratio of neutral gas to blue
luminosity of $0.6$ (Schmidt \& Boller, 1992a,b), also rather normal for
its type. As this dwarf galaxy appears relatively regular on survey
plates and was indicated to be nearby, I selected it for a kinematic
study of its stars and its ionized gas as well as for a study of its
chemical abundances through optical long slit spectroscopy.

Only a radial velocity exists as distance indicator for UGC
685. Surface photometry which is necessary for the analysis of
kinematic data is still missing. Finally, no colors or other values
which hint at the star formation history, are published yet. I
therefore included this dwarf galaxy in photometric CCD observations
with the Calar Alto 3.5m and 2.2m telescopes to obtain multi-color
surface photometry of the galaxy, its structural parameters and HII
morphology and to resolve the brightest supergiants. These can yield an
independent distance estimate, following the recent update of the
calibration for the brightest blue supergiants by K\&T and R\&RR.
The limitation and error budget of this techniques has been discussed
in detail by R\&RR.

The observations and their reductions are described in section 2. The
structure and HII morphology, the color-magnitude diagram of the
resolved stars and the distance estimate are presented in section 3
and I conclude in section 4.

\section{Observations and Reductions}

\subsection{B and R band observations}

Prime focus CCD images were obtained with the Calar Alto 3.5 m
telescope during a photometric night, the details are given
in table 1.
After debiasing and flatfielding in the usual manner, I constructed an
image which contains only the underlying smooth light distribution of
the unresolved fainter stars as described by Hopp \& Schulte-Ladbeck (1991). 
This 'smooth mask' was used to derive the structural parameters by applying
the ellipticity fit of Bender \& M\"ollenhoff (1987). Then, the 'smooth
masks' were subtracted from the original B and R images which yielded
frames with the resolved stars only. To these frames, DAOPHOT in its MIDAS
version was applied, again following Hopp \& Schulte-Ladbeck (1991). The
frames of both colors were searched independently by DAOPHOT and only those
objects were accepted, which were found in both frames.
Figure 1 shows the B image of UGC 685.

\subsection{H${\alpha}$ observations}

The Calar Alto 2.2m telescope and its CCD camera were used to obtain
H${\alpha}$- and R-images of UGC 685. The night was
of poor photometric quality, thus only a preliminary calibration
through the R band observations was applied. Details are given in
table 1. As above, the usual CCD calibration frames were obtained and applied
and the individual exposures in each filter were added.  The R image
was flux-scaled to the H${\alpha}$-image with the aid of several
medium-bright stars and then subtracted, yielding a continuum free
H${\alpha}$ frame of UGC 685 (for more technical details see Bomans
et al, 1997). Figure 2 shows the H${\alpha}$ image of UGC 685.

\subsection{ J band observations}

J-band observations were taken with the Calar Alto 3.5m telescope and
its IR camera MAGIC during a photometric night. The set-up details can be
found in table 1. After every five object frames, nearby night sky
data were taken at 4 different positions around UGC 685. The sets of
object frames were slightly shifted against each other. The
individual 10 sec. exposures were corrected for bias, dark current,
night sky pattern and fixed-pattern noise with the usual calibration
data. Then they were transformed into a common reference frame and
finally added. 

As the on-line facility did not show any signal of UGC
685, no K data were obtained. For the same reason, centering was not
perfect and the eastern outskirt of UGC 685 was therefore not mapped
within the tiny field of view available with this set-up.

\begin{table*}
\caption{Details of the observations presented in this paper.}
\begin{tabular}{l|c|c|c}
\hline
observations & B,R$_{(c)}$ & H${\alpha}$,R &  J \\
\hline
date                  & Oct 12, 1994  & Oct 04, 1994  & Sep. 06, 1995 \\
telescope             & CA 3.5m Prim  & CA 2.2m Cas   & 3.5 m Cas \\
detector              & Tektronix CCD & Tektronix CCD & NICMOS3 HgCdTe \\
pixels                & 1024 x 1024   & 1024 x 1024   & 256 x 256 \\
pixel size [\arcsec]  & 0.53      &  0.28       &  0.33 \\
field of view [\arcmin] & 7 & 4.7 & 1.4 \\
                      & (circle)  & (square)    & (square) \\
total exp. time [sec] & 360, 240  & 900, 680    & 1200 \\
seeing [FWHM \arcsec] & 1.1, 1.3  &  1.7, 1.7   & 1.5 \\
standard stars        & NGC 2419, NGC 7790 & -  & Elias et al \\
                      & Christian et al, 1985 & & 1982\\
\hline
\end{tabular}
\end{table*}

\begin{table*}
\caption{
Overall properties of UGC 685 from the literature and as
derived in this paper. Absolute magnitudes are corrected for galactic
reddening with the given extinction value from Burstein \& Heiles (1982),
distance depend values are converted with the new distance estimate of
5.5 Mpc.}
\begin{tabular}{l|c|l}
\hline
property                   & value           & reference\\
\hline
RA (1950)                  & 01 04 43        & NED \\
Decl. (1950)               & +16 25 01       & NED \\
Galactic latitude          & -46.02          & UGC \\
E$_{B-V}$                  & 0.10            & Burstein \& Heiles 1982\\
distance                   & 0.27 D$_{Virgo}$& Kran-Korteweg 1986\\
                           &$\sim$ 6 Mpc     & \\
distance [Mpc]             & 5.5 $\pm$ 30\%  & this paper\\
HI $S_{\nu}d\nu$ [Jy km/s] & 12.6            & Lu et al 1993 \\
HI velocity  [km/s]        & 157 $\pm$ 1     & Lu et al 1993 \\
HI W50 width [km/s]        & 73 $\pm$ 1      & Lu et al 1993 \\
HI mass      [\Msolar]     & 8.5 10$^7$      & Hoffman et al 1996 \\
V$_{max}$    [km/s]        & 75              & Hoffman et al, 1996\\
FIR Iras flux              & below limits    & Schmidt \& Boller 1992a\\
                           &                 & \\
B                          & 14.55 $\pm$ 0.08      & this paper \\
$M_{B,25.0}$               & -14.48                & this paper \\
$M_{B,asym}$               & -14.92                & this paper \\
B-R                        & +0.97 $\pm$ 0.05      & this paper\\
B-I                        & +1.39                 & Lu et al 1993 and here\\
B-J                        & +1.55 $\pm$ 0.15      & this paper\\
$\mu_{0,B}$ [mag/$\Box$\arcsec] & 21.57 $\pm$ 0.09 & this paper\\
$\mu_{0,R}$ [mag/$\Box$\arcsec] & 20.65 $\pm$ 0.06 & this paper\\
$\mu_{0,J}$ [mag/$\Box$\arcsec] & 20.11 $\pm$ 0.10 & this paper\\
scale length B [pc]       & 338 $\pm$ 16          & this paper\\
scale length R [pc]       & 330 $\pm$ 26          & this paper\\
scale length J [pc]       & 338 $\pm$ 21          & this paper\\
a$_{26.5 mag/\Box\arcsec}$ & 55'' (1.46 kpc)       & this paper\\
Elipticity                 & 0.33                  & this paper \\
Number of HII regions      & 5                     & this paper\\
lg L$_{H\alpha}$ [erg s$^{-1}$] & 38.6             & this paper\\
\hline
\end{tabular}
\end{table*}


\begin{figure} 
\centerline{\psfig{figure=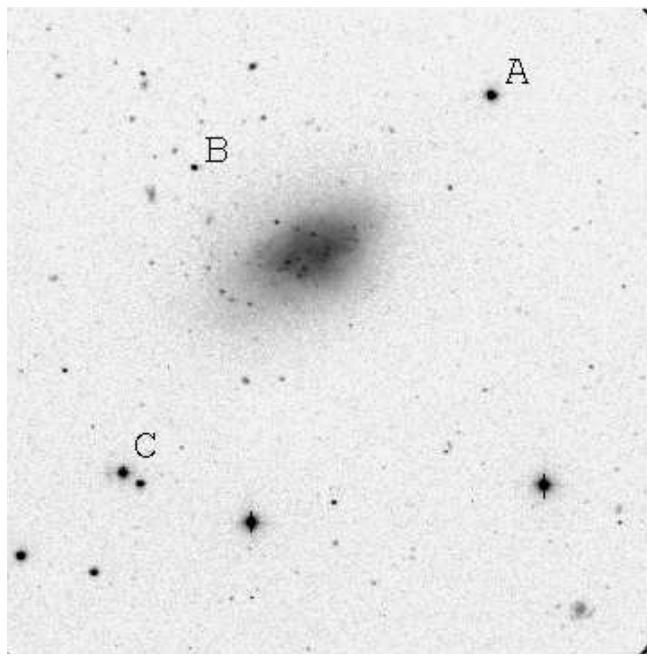,width=8.5cm,clip=t}}
\caption[]{CCD 3.5m telescope B image of UGC 685. North top, East
left. The {\bf horizontal vertical?} 
side length is 346\arcsec. The letters identify
those very bright stars where residuals are still visible in Fig. 2.}
\end{figure}

\begin{figure} 
\centerline{\psfig{figure=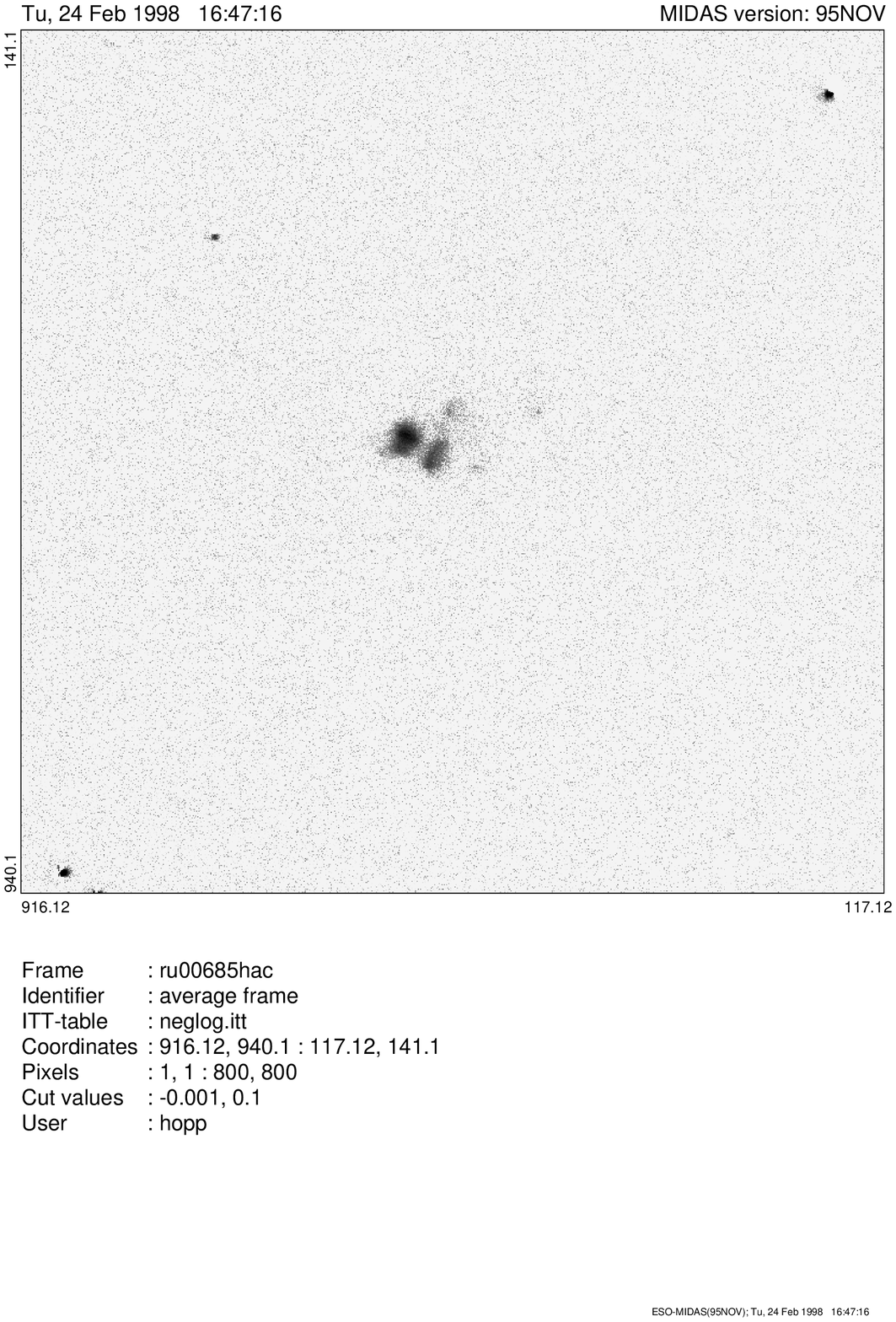,width=8.5cm,clip=t}}
\caption[]{CCD 2.2m telescope image in H${\alpha}$ of UGC 685
after continuum subtraction. North top, East
left. The side length is 160\arcsec. The residuals seen at large
distances from UGC 685 are from saturated stars which are identified
in Fig. 1 by letters.}
\end{figure}

\begin{figure} \centerline{\psfig{figure=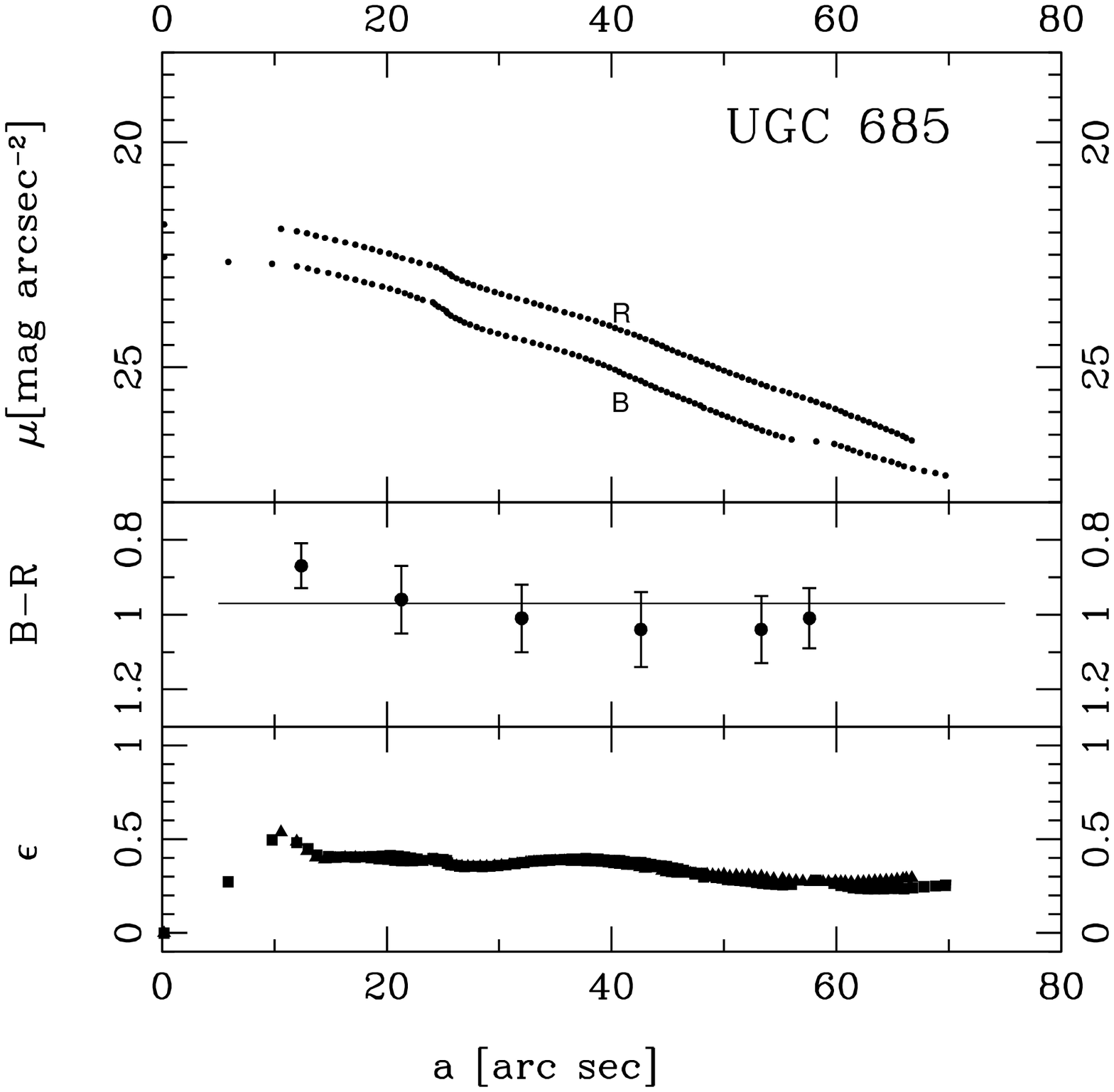,width=8.5cm,angle=0}}
\caption[]{Top: The surface brightness profiles according to the
ellipticity fits in the red (upper) and blue (lower) 3.5m CCD frames.
Middle: The color gradient as calculated from the curves in the upper
graph. The straight line gives the color for the total light.  Bottom:
The radial ellipticity variation derived from the blue (squares) and red
(triangles) frames (all 3.5m telescope CCD data).}
\end{figure}

\section{Results and Discussion}

\subsection{Morphology and total fluxes}

The B (Fig. 1), R, and J images show a mostly regular galaxy of
elliptical shape. Only few brighter and extended knots in a region to
the South-East of the center can be seen.  The H${\alpha}$ (Fig. 2)
image reveals that these knots are HII regions. To the limits of the
images, one bright and 4 fainter, smaller HII regions were identified.
The largest one has a diameter of about 9.3\arcsec (0.25 kpc). These
few HII regions are clumped in an area of no more than 25\arcsec
(0.65 kpc) diameter. This blob of HII regions is off-set from the
center by 11\arcsec (0.31 kpc) or about one scale length of the
stellar light distribution.  Very little diffuse H${\alpha}$ emission
outside the HII regions is visible, but the image which was taken for
selecting proper long slit settings might have an exposure time which
was too short to detect very faint structures like those describes for
example in the case of NGC 4449 by Bomans et al (1997). The long
slit spectra along the major axis of UGC 685 cross some of the
identified HII regions and indeed reveal the typical emission line
pattern of HII regions. Outside these HII regions, but still well inside 
the low surface brightness envelope, sources with the shape of the
point-spread function are visible. These objects trace the resolved
brightest stars - mostly supergiants - in UGC 685 even though it is easily
possible that some of these 'stars' may be in reality composed objects
like blends of stars or stellar clusters.

A mask was constructed at the 25.0 mag/$\Box$\arcsec ~level in B and
the flux in B and R inside this isophote was determined, yielding B =
14.55 and B-R = + 0.97 (table 2). Errors are less than 0.1 and are
mostly of systematic nature. The B magnitude is in good agreement with
those in the literature (see Schmidt \& Boller, 1992a). The absolute
values derived from these observed magnitudes were computed using the distance
estimate of 5.5 Mpc from chapter 3.4 and a correction for galactic foreground
extinction as given by Burstein \& Heiles (1982). No attempt
was done to correct for internal reddening. As already mentioned, the
J frames exclude a small part of the faint eastern outskirt, less
than 4\% of the total area. These outskirts contribute little light,
nevertheless the derived J magnitude of 13.00 is more like a good lower
limit to the total flux.

I measured the flux in the continuum-free H${\alpha}$ image and found
a flux of $1~10^{-13}~erg~s^{-1}~cm^{-2}$ which converts to an
absolute flux in the H${\alpha}$ line as given in table 2 (using D =
5.5 Mpc). The Poisson error of this value
is small, less than 1\%, but almost negligible compared to
the error of the zero point calibration of about 20\%.  Hunter et al.
(1994) observed H${\alpha}$ line fluxes for a sample of late-type
galaxies, mostly dwarf irregulars. In comparison to this sample, the
H${\alpha}$ line flux of UGC 685 is normal for an irregular dwarf of
its total magnitude and indicates a relatively low recent star
formation rate (see below).

\subsection{Surface photometry}

Figure 3 shows the surface brightness profile, the radial color profil
and the variation of the ellipticity with major axis for B and R. UGC
685 has a rather elliptical shape with a mean ellipticity value of
0.33 for both colors.  The ellipticity varies only slightly with
radius while the (not shown) position angle increases steadily by the
small, but significant amount of 20 degree from 10 to 70\arcsec. The
surface brightness profile can be traced out to about 70\arcsec~in B
and 65\arcsec~in R. It is interesting to note that in neutral
hydrogen, the profile of UGC 685 was traced to a 2.6 times larger
radius than here in the optical (see Hoffman et al, 1996 for the HI data).
In J, the galaxy was traced down to 23.1 mag/$\Box$\arcsec and to a
distance of 35\arcsec. Within this more limited range, the
ellipticity is essentially the same (0.32) as in the CCD color bands.

In all three colors, the profile can be well described by an exponential law
outside a central region of about 10\arcsec. The central profiles are
significantly flatter. One should remember that in the very center,
several knots (HII regions, stars, clusters) were removed, therefore it
is difficult to trace the profile there. Outside 10\arcsec, I fitted
an exponential brightness profile to the data and got a scale length of
12.4\arcsec to 12.7\arcsec (table 2). Thus, the scale length of the stellar
distribution is 10 times smaller than the one of the HI distribution
(Hoffman et al, 1996). The obtained parameters, already converted to
metric values with the distance estimated below, are given in table
2. They are quite normal for a galaxy of this size and
luminosity\footnote{The R band data from the 2.2m
telescope yielded very similar results: The profile is detected out to
about 55\arcsec, is flat in the center, and shows an exponential
distribution law with a scale length of 16 $\pm$ 1.2\arcsec.}. The
central surface brightnesses are relatively high for an irregular dwarf
with only moderate signs of recent star formation. A color
profile was calculated (Fig. 3). Most of the body has the same color
while the central part is slightly bluer. B-J, albeit having a larger
systematic error, shows no significant variation with radius.

The exponential law parameters were used to calculate asymptotic
total magnitudes of 14.11, 13.14, and 12.60 in B, R, and J, respectively.

H${\alpha}$ emission was only detected in a small area near to the
center of UGC 685, slightly off towards the south-east. Thus, the
recent places of star formation activity show a strong asymmetry in
the angular distribution and are well localized. I measured the
H${\alpha}$ surface brightness as a function of radius and averaged
over the azimuth (for a better comparison with similar profiles for
other irregular galaxies presented by Hunter et al, 1998). Naturally,
this radial H${\alpha}$ profile is only valid for the small sector
(Fig. 2). Outside, the H${\alpha}$ surface brightness is below the
detection limits (about $1.5~10^{-16}~erg~s^{-1}~cm^{-2}$). Assume
that the H${\alpha}$ flux is a good measure of the total number of
ionizing photons emitted in a galaxy. This number can be compared to
the photons expected from massive stars for a given initial mass
function (IMF). Taking into account the lifetimes of the massive
stars, a formation rate of hot stars can be derived and extended to a
total star formation rate (SFR) by extrapolating to all
stars with the IMF. Hunter \& Gallagher (1986, see also Gallagher et
al., 1984) used a Salpeter
function to establish a conversion formula which I used to transform
the observed H${\alpha}$ surface brightness of UGC 685 into a star
formation rate per ${\rm pc}^{2}$. A distance of 5.5 Mpc was used. Fig. 4
show the radial profile. Given the uncertainty in the H${\alpha}$ flux
calibration, the only secure feature is the off-center peak in the
recent specific star formation activity, and an overall trend that the
recent star formation activity drops to larger radii. The shape of the
radial distribution of the recent star formation activity in UGC 685
is similar to the one observed in Sex A and IC 1613 (Hunter et al, 1998),
even though its overall value is lower than in those two dwarfs. As discussed
in more detail by van Zee et al (1998), an azimutal average of the
local SFR (as measured by the H${\alpha}$ flux) can be
misleading for irregular dwarf galaxies where the star formation often
takes place at only a few places. UGC 685 is obviously a good example
for this case. Using again the derivation of Hunter \& Gallagher
(1986), the total H${\alpha}$ flux from chapter 3.1 can be transformed
into an average SFR over the last $\sim 10^7$~yr. I
derived 0.003 \Msolar yr$^{-1}$ with an error based on the measurement
alone of about 20\%. Here again, the result depends on the applied
distance value of 5.5 Mpc. 

\begin{figure} 
\centerline{\psfig{figure=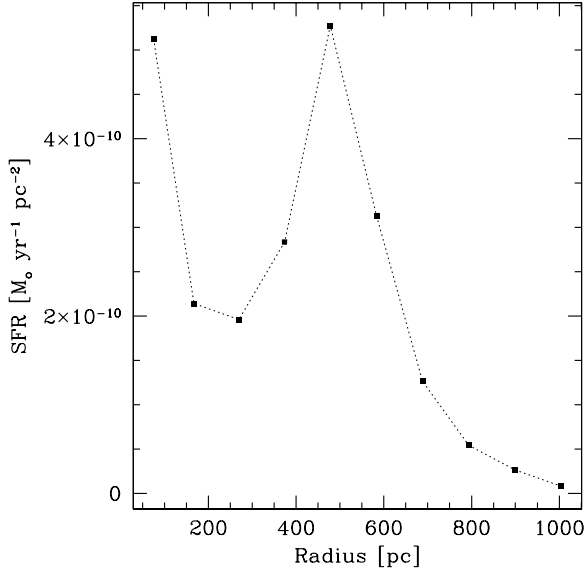,width=8.5cm,clip=t}}
\caption[]{The radial distribution of the recent star formation activity as 
derived from the  H${\alpha}$ flux applying the transformation
relation between flux and star formation rate as given by Hunter \&
Gallagher (1986). A distance of 5.5 Mpc was assumed. 
This curve is only valid for the sector where HII regions are visible, 
outside this sector the star formation rate is below the detection limit.}
\end{figure}

The radial color gradient shows no indication that the average SFR
varies strongly with location (Fig. 3). A B-R color
map shows a quite homogeneous distribution with only small (5\%
level) deviations in the central area, where recent (and localized)
star formation is already indicated by the HII regions.
 
\subsection{Resolved stellar population}

A total of 209 stars are detected in both (B and R) frames, down to
a limiting magnitude of 24 in both filters. These stars are identified
in table 3\footnote{An upper flux limit had been set when running
DAOPHOT to avoid saturated stars. Therefore, the brightest stars
visible in fig. 1 can not be found in table 3, but they are 
Milky Way objects anyway.}.  Fainter than 22.5 in R and 23.0 in B, the
photometric errors as estimated by the DAOPHOT PSF fitting algorithm
are larger than 0.1. 73 of these stars are projected on or belong to
UGC 685. As the CCD frames are by far larger than the galaxy, I can
determine the surface density of fore- and background objects with
stellar PSF. A statistical correction of 5.1 stars was calculated which
leaves about 68 stars belonging to UGC 685. The color distribution of
the field stars outside UGC 685 (see Fig. 5) shows that most of the
contamination contributes in the red while practically all detected
blue objects (B-R $<$ 0.5) belong to the young population in
UGC 685.

Artificial star tests were applied inside DAOPHOT to estimate the
completeness of the star detection especially in the magnitude range
of the supergiants of UGC 685. 50 stars were added in 1 magnitude bins
in several experiments for each of these bins.  The recovery rate
indicates that the completeness starts to drop from almost 100 \% at
21.9 (22.7) in R (B) and the 50\% completeness magnitudes are about
22.9 (23.7) in R (B).

\setcounter{table}{2}
\begin{table}
\caption{Stars found in B and R on the UGC 685 frames. Given are
a running number (from B frame), relative frame coordinates in arc sec
where +x increases towards East and +y increases towards South, the B 
magnitudes and their errors according to DAOPHOT (which does not include
systematic errors of the calibrations) followed by the same photometric
informations for R, and finally B-R.}  
\begin{tabular}{rrrrrrrr}
\hline
 No.  &   X  &  Y  &  B  &  $\sigma_B$ &  R  &  $\sigma_R$  & B-R\\
~& arc sec & arc sec & mag & mag & mag & mag & mag \\
\hline
  12 & 323.7  &  11.2 &  22.19  &  0.05 &  21.08 &   0.06 &  1.11 \\
  13 & 280.5  &  11.6 &  22.85  &  0.05 &  20.63 &   0.08 &  2.22 \\
  18 & 106.1  &  16.4 &  23.77  &  0.21 &  22.42 &   0.11 &  1.35 \\
  29 & 221.1  &  28.6 &  23.58  &  0.11 &  21.39 &   0.07 &  2.19 \\
  31 & 287.6  &  29.6 &  21.94  &  0.05 &  20.22 &   0.04 &  1.73 \\
  32 & 213.5  &  31.6 &  20.14  &  0.05 &  18.74 &   0.05 &  1.40 \\
  35 &   3.4  &  32.9 &  23.38  &  0.13 &  21.54 &   0.09 &  1.84 \\
  39 & 272.6  &  35.7 &  20.28  &  0.02 &  18.91 &   0.02 &  1.37 \\
  40 & 318.5  &  36.9 &  21.67  &  0.04 &  20.08 &   0.05 &  1.59 \\
  45 & 271.9  &  41.5 &  21.72  &  0.07 &  20.78 &   0.09 &  0.93 \\
  52 &   2.0  &  53.1 &  22.80  &  0.09 &  20.87 &   0.08 &  1.93 \\
  53 & 108.7  &  56.7 &  23.52  &  0.11 &  21.99 &   0.07 &  1.53 \\
  55 & 181.2  &  60.1 &  23.42  &  0.16 &  21.50 &   0.07 &  1.91 \\
  56 & 207.6  &  59.7 &  21.30  &  0.03 &  20.39 &   0.03 &  0.91 \\
  58 & 278.5  &  60.1 &  21.83  &  0.04 &  20.81 &   0.03 &  1.02 \\
  59 & 218.1  &  61.5 &  22.72  &  0.07 &  21.10 &   0.07 &  1.62 \\
  60 & 318.8  &  61.0 &  23.80  &  0.23 &  21.67 &   0.09 &  2.13 \\
  62 & 241.1  &  61.8 &  22.34  &  0.07 &  21.33 &   0.07 &  1.01 \\
  66 & 178.0  &  65.5 &  22.22  &  0.05 &  21.86 &   0.07 &  0.37 \\
  68 & 304.9  &  67.1 &  23.35  &  0.14 &  21.78 &   0.09 &  1.56 \\
  67 & 138.3  &  66.9 &  22.55  &  0.06 &  21.15 &   0.09 &  1.40 \\
  69 &   6.2  &  67.9 &  22.26  &  0.05 &  20.60 &   0.06 &  1.66 \\
  79 & 158.0  &  75.4 &  23.31  &  0.14 &  21.86 &   0.12 &  1.45 \\
  81 & 160.3  &  77.4 &  23.32  &  0.13 &  22.79 &   0.18 &  0.53 \\
  82 &  57.8  &  78.1 &  23.06  &  0.07 &  21.59 &   0.09 &  1.47 \\
  83 & 255.6  &  77.6 &  21.66  &  0.07 &  20.88 &   0.09 &  0.77 \\
  87 & 266.5  &  80.8 &  22.45  &  0.04 &  20.72 &   0.03 &  1.73 \\
  91 & 244.8  &  86.7 &  18.73  &  0.02 &  17.16 &   0.03 &  1.58 \\
  95 & 199.8  &  91.2 &  23.25  &  0.11 &  21.66 &   0.08 &  1.59 \\
  96 &  26.5  &  94.2 &  22.69  &  0.07 &  22.00 &   0.10 &  0.70 \\
  97 & 165.2  &  94.5 &  22.67  &  0.10 &  20.87 &   0.06 &  1.81 \\
  98 & 106.7  &  97.8 &  20.94  &  0.02 &  20.06 &   0.03 &  0.88 \\
  99 & 270.5  &  98.6 &  22.52  &  0.09 &  21.27 &   0.09 &  1.26 \\
 104 & 267.9  & 101.4 &  21.62  &  0.09 &  19.74 &   0.08 &  1.87 \\
 111 & 222.3  & 107.0 &  22.81  &  0.11 &  20.59 &   0.03 &  2.22 \\
 113 & 246.3  & 108.1 &  24.09  &  0.19 &  22.98 &   0.18 &  1.11 \\
 114 & 324.0  & 107.9 &  23.42  &  0.11 &  22.14 &   0.10 &  1.28 \\
 123 & 236.5  & 114.3 &  22.45  &  0.08 &  20.25 &   0.06 &  2.20 \\
 124 & 157.4  & 114.8 &  23.50  &  0.15 &  22.27 &   0.11 &  1.22 \\
 128 &  31.1  & 116.5 &  22.68  &  0.07 &  20.82 &   0.05 &  1.86 \\
 133 & 200.5  & 116.4 &  20.22  &  0.02 &  19.29 &   0.02 &  0.94 \\
 134 & 237.3  & 116.1 &  22.21  &  0.05 &  20.06 &   0.06 &  2.15 \\
 135 & 144.2  & 117.6 &  24.13  &  0.28 &  23.27 &   0.24 &  0.86 \\
 138 & 149.4  & 118.4 &  22.68  &  0.06 &  21.48 &   0.07 &  1.20 \\
 143 & 158.7  & 119.4 &  21.66  &  0.04 &  20.97 &   0.03 &  0.70 \\
 145 &  39.3  & 120.2 &  22.80  &  0.09 &  21.87 &   0.10 &  0.93 \\
 146 & 189.2  & 120.3 &  21.11  &  0.03 &  20.65 &   0.05 &  0.46 \\
 149 & 165.2  & 121.4 &  21.77  &  0.04 &  21.43 &   0.05 &  0.34 \\
 157 & 184.6  & 123.9 &  22.71  &  0.09 &  22.57 &   0.20 &  0.14 \\
 155 & 180.9  & 123.1 &  20.54  &  0.02 &  19.61 &   0.01 &  0.92 \\
 165 & 158.8  & 125.9 &  22.05  &  0.04 &  21.39 &   0.06 &  0.66 \\
 166 & 161.9  & 125.5 &  22.00  &  0.08 &  21.62 &   0.11 &  0.37 \\
 169 & 173.7  & 126.2 &  21.91  &  0.04 &  21.62 &   0.08 &  0.30 \\
 171 & 156.7  & 126.7 &  22.15  &  0.06 &  22.26 &   0.12 & -0.11 \\
 176 & 171.9  & 128.7 &  22.18  &  0.08 &  20.82 &   0.06 &  1.36 \\
\hline							    
\end{tabular}						    
\end{table}

\setcounter{table}{2}
\begin{table}
\caption{Table 3 continued }
\begin{tabular}{rrrrrrrr}
\hline
 No.  &   X  &  Y  &  B  &  $\sigma_B$ &  R  &  $\sigma_R$  & B-R\\
~& arc sec & arc sec & mag & mag & mag & mag & mag \\
\hline
 180 &  68.5  & 129.3 &  23.25  &  0.13 &  21.81 &   0.06 &  1.44 \\
 186 & 167.9  & 130.2 &  22.43  &  0.06 &  21.69 &   0.10 &  0.74 \\
 189 & 173.9  & 130.7 &  21.77  &  0.04 &  21.24 &   0.07 &  0.52 \\
 193 & 162.6  & 130.9 &  22.46  &  0.07 &  21.61 &   0.11 &  0.85 \\
 194 & 172.4  & 131.3 &  21.42  &  0.05 &  21.27 &   0.04 &  0.16 \\
 195 & 189.2  & 131.0 &  23.31  &  0.10 &  22.35 &   0.14 &  0.97 \\
 198 & 209.9  & 132.6 &  22.29  &  0.06 &  21.88 &   0.07 &  0.41 \\
 200 & 179.2  & 133.2 &  22.34  &  0.06 &  21.55 &   0.06 &  0.79 \\
 203 & 165.0  & 133.5 &  24.14  &  0.29 &  22.11 &   0.16 &  2.02 \\
 206 & 222.8  & 133.6 &  22.03  &  0.04 &  21.36 &   0.04 &  0.67 \\
 207 & 173.8  & 134.1 &  20.67  &  0.03 &  20.15 &   0.03 &  0.52 \\
 208 & 180.1  & 134.3 &  22.51  &  0.07 &  21.12 &   0.05 &  1.39 \\
 212 & 195.6  & 135.5 &  23.11  &  0.14 &  22.59 &   0.17 &  0.51 \\
 214 & 209.4  & 134.7 &  22.66  &  0.11 &  22.00 &   0.10 &  0.66 \\
 216 & 188.9  & 136.3 &  21.23  &  0.02 &  19.55 &   0.02 &  1.68 \\
 217 & 175.7  & 136.8 &  21.73  &  0.04 &  21.32 &   0.06 &  0.40 \\
 221 & 182.1  & 137.7 &  23.54  &  0.22 &  22.14 &   0.15 &  1.40 \\
 222 & 204.4  & 138.2 &  22.48  &  0.08 &  21.29 &   0.05 &  1.19 \\
 223 &  92.9  & 138.4 &  23.38  &  0.13 &  21.65 &   0.05 &  1.73 \\
 226 & 251.9  & 138.7 &  23.37  &  0.12 &  21.64 &   0.05 &  1.73 \\
 227 & 268.1  & 138.7 &  21.90  &  0.05 &  21.03 &   0.07 &  0.87 \\
 229 & 215.3  & 138.9 &  22.04  &  0.07 &  21.98 &   0.06 &  0.06 \\
 232 & 209.5  & 140.2 &  21.85  &  0.04 &  21.49 &   0.04 &  0.35 \\
 233 & 176.2  & 140.8 &  22.48  &  0.06 &  22.71 &   0.20 & -0.22 \\
 238 & 198.6  & 142.4 &  21.79  &  0.05 &  21.36 &   0.05 &  0.43 \\
 237 & 237.3  & 141.6 &  21.67  &  0.03 &  20.82 &   0.04 &  0.85 \\
 239 & 205.4  & 142.0 &  22.91  &  0.14 &  22.30 &   0.15 &  0.61 \\
 241 & 200.7  & 142.6 &  22.67  &  0.09 &  22.32 &   0.10 &  0.34 \\
 243 & 173.0  & 143.4 &  23.21  &  0.15 &  22.94 &   0.24 &  0.27 \\
 247 & 177.5  & 143.7 &  23.71  &  0.18 &  22.92 &   0.29 &  0.80 \\
 245 & 193.0  & 143.4 &  23.27  &  0.29 &  22.32 &   0.13 &  0.95 \\
 246 & 168.9  & 143.8 &  22.92  &  0.13 &  21.83 &   0.10 &  1.09 \\
 249 & 249.6  & 145.1 &  22.45  &  0.06 &  21.81 &   0.08 &  0.64 \\
 250 & 188.3  & 145.9 &  21.10  &  0.05 &  21.06 &   0.07 &  0.04 \\
 262 & 320.8  & 147.2 &  22.05  &  0.07 &  20.59 &   0.07 &  1.46 \\
 263 & 180.0  & 148.7 &  23.52  &  0.14 &  22.39 &   0.16 &  1.13 \\
 265 & 203.5  & 150.1 &  22.75  &  0.09 &  22.23 &   0.08 &  0.52 \\
 269 & 178.6  & 151.0 &  22.08  &  0.05 &  21.77 &   0.09 &  0.31 \\
 270 &  12.9  & 151.3 &  21.77  &  0.04 &  19.92 &   0.03 &  1.85 \\
 271 & 207.0  & 151.4 &  23.99  &  0.21 &  22.20 &   0.08 &  1.79 \\
 273 & 149.0  & 153.7 &  23.10  &  0.10 &  22.23 &   0.10 &  0.88 \\
 274 & 230.0  & 154.2 &  20.90  &  0.03 &  19.96 &   0.03 &  0.95 \\
 279 & 224.9  & 158.8 &  22.08  &  0.06 &  20.68 &   0.10 &  1.40 \\
 282 & 215.1  & 161.3 &  20.97  &  0.02 &  20.02 &   0.02 &  0.95 \\
 286 &  69.3  & 164.8 &  23.24  &  0.12 &  21.80 &   0.10 &  1.44 \\
 290 &  65.5  & 166.3 &  23.35  &  0.12 &  21.87 &   0.08 &  1.49 \\
 293 & 245.5  & 167.6 &  23.88  &  0.21 &  22.79 &   0.18 &  1.09 \\
 297 & 160.4  & 172.7 &  21.86  &  0.04 &  20.82 &   0.04 &  1.03 \\
 302 & 247.2  & 175.9 &  22.45  &  0.06 &  20.63 &   0.03 &  1.82 \\
 304 & 276.8  & 179.4 &  23.61  &  0.14 &  22.57 &   0.14 &  1.03 \\
 306 & 211.7  & 180.1 &  22.79  &  0.07 &  21.51 &   0.05 &  1.28 \\
 310 & 246.9  & 185.4 &  24.06  &  0.22 &  22.72 &   0.13 &  1.34 \\
 313 & 330.5  & 189.1 &  22.97  &  0.09 &  22.03 &   0.10 &  0.94 \\
 317 &  30.8  & 195.2 &  21.58  &  0.03 &  19.87 &   0.04 &  1.72 \\
 320 & 315.0  & 197.2 &  19.56  &  0.03 &  17.75 &   0.06 &  1.81 \\
 323 & 338.5  & 197.5 &  22.46  &  0.06 &  20.74 &   0.04 &  1.71 \\
 325 &   2.6  & 201.0 &  23.17  &  0.08 &  22.25 &   0.14 &  0.91 \\
 327 & 197.8  & 202.1 &  21.80  &  0.03 &  21.05 &   0.03 &  0.76 \\
 328 &  26.0  & 202.7 &  23.35  &  0.08 &  21.89 &   0.07 &  1.46 \\
 331 & 216.9  & 203.0 &  21.25  &  0.06 &  20.36 &   0.08 &  0.89 \\
 336 &  33.7  & 207.1 &  23.44  &  0.12 &  22.11 &   0.08 &  1.33 \\
 337 &  88.0  & 207.9 &  21.55  &  0.05 &  19.70 &   0.04 &  1.84 \\
\hline							    
\end{tabular}						    
\end{table}

\setcounter{table}{2}
\begin{table}
\caption{Table 3 continued.}
\begin{tabular}{rrrrrrrr}
\hline
 No.  &   X  &  Y  &  B  &  $\sigma_B$ &  R  &  $\sigma_R$  & B-R\\
~& arc sec & arc sec & mag & mag & mag & mag & mag \\
\hline
 340 & 300.9  & 210.1 &  23.89  &  0.20 &  21.80 &   0.08 &  2.09 \\
 342 & 125.2  & 212.0 &  23.11  &  0.08 &  21.87 &   0.09 &  1.24 \\
 343 &  44.4  & 213.9 &  23.39  &  0.09 &  21.93 &   0.11 &  1.46 \\
 345 & 273.3  & 216.0 &  23.05  &  0.06 &  21.47 &   0.08 &  1.58 \\
 349 & 178.3  & 220.8 &  23.41  &  0.07 &  21.95 &   0.09 &  1.46 \\
 351 & 134.5  & 224.3 &  24.10  &  0.23 &  22.85 &   0.19 &  1.25 \\
 353 & 300.2  & 227.9 &  21.99  &  0.05 &  20.98 &   0.08 &  1.01 \\
 354 &  56.7  & 229.2 &  23.39  &  0.09 &  22.01 &   0.09 &  1.38 \\
 356 & 102.1  & 231.0 &  22.67  &  0.06 &  21.96 &   0.08 &  0.70 \\
 357 &   6.9  & 232.2 &  22.76  &  0.09 &  21.65 &   0.07 &  1.10 \\
 358 &  24.5  & 232.2 &  23.80  &  0.21 &  21.63 &   0.07 &  2.18 \\
 361 & 199.6  & 234.5 &  23.72  &  0.17 &  22.50 &   0.12 &  1.22 \\
 366 & 107.5  & 237.5 &  22.19  &  0.08 &  21.35 &   0.06 &  0.84 \\
 372 & 251.9  & 239.4 &  22.32  &  0.07 &  21.42 &   0.07 &  0.91 \\
 375 & 109.1  & 241.6 &  20.93  &  0.05 &  19.31 &   0.05 &  1.62 \\
 380 & 233.7  & 247.2 &  23.03  &  0.09 &  21.59 &   0.07 &  1.44 \\
 384 & 249.6  & 247.9 &  24.72  &  0.30 &  23.40 &   0.26 &  1.32 \\
 387 & 303.2  & 248.1 &  23.48  &  0.13 &  22.13 &   0.12 &  1.35 \\
 385 &  89.2  & 248.2 &  22.22  &  0.07 &  21.27 &   0.07 &  0.96 \\
 391 &  16.9  & 251.3 &  23.06  &  0.11 &  21.79 &   0.10 &  1.27 \\
 396 & 262.6  & 254.3 &  23.64  &  0.16 &  22.20 &   0.07 &  1.44 \\
 400 & 166.7  & 261.7 &  23.57  &  0.14 &  21.79 &   0.06 &  1.79 \\
 402 &   1.1  & 262.8 &  23.24  &  0.12 &  22.64 &   0.14 &  0.61 \\
 405 & 227.4  & 264.3 &  23.45  &  0.15 &  22.37 &   0.13 &  1.09 \\
 407 & 169.6  & 269.2 &  19.40  &  0.04 &  17.72 &   0.05 &  1.68 \\
 409 & 128.4  & 271.3 &  23.47  &  0.13 &  21.27 &   0.06 &  2.21 \\
 408 &  62.0  & 271.6 &  23.25  &  0.09 &  22.38 &   0.14 &  0.87 \\
 410 &  15.9  & 272.9 &  22.40  &  0.08 &  20.66 &   0.07 &  1.74 \\
 417 &  15.1  & 280.2 &  20.44  &  0.05 &  18.80 &   0.07 &  1.65 \\
 418 & 277.1  & 282.8 &  23.59  &  0.16 &  22.20 &   0.09 &  1.39 \\
 421 &  82.3  & 285.4 &  23.57  &  0.16 &  22.02 &   0.08 &  1.56 \\
 424 & 314.4  & 287.2 &  23.74  &  0.17 &  22.54 &   0.15 &  1.20 \\
 427 &  52.8  & 290.5 &  22.80  &  0.08 &  21.65 &   0.11 &  1.15 \\
 428 & 168.9  & 291.4 &  21.66  &  0.05 &  20.60 &   0.07 &  1.07 \\
 430 &  43.8  & 292.8 &  23.52  &  0.12 &  22.66 &   0.16 &  0.85 \\
 432 & 126.9  & 294.7 &  23.26  &  0.10 &  21.77 &   0.08 &  1.49 \\
 433 & 215.1  & 294.8 &  22.93  &  0.05 &  22.48 &   0.14 &  0.45 \\
 436 &  36.8  & 297.6 &  23.58  &  0.14 &  21.53 &   0.09 &  2.04 \\
 440 & 127.7  & 299.4 &  23.02  &  0.09 &  22.10 &   0.08 &  0.92 \\
 444 & 322.7  & 303.2 &  23.92  &  0.25 &  22.53 &   0.14 &  1.39 \\
 446 & 299.2  & 307.4 &  16.62  &  0.05 &  15.45 &   0.04 &  1.17 \\
 445 & 190.3  & 307.2 &  23.20  &  0.09 &  22.22 &   0.13 &  0.98 \\
 447 & 346.9  & 309.9 &  23.24  &  0.15 &  21.68 &   0.09 &  1.55 \\
 451 & 148.0  & 313.2 &  22.37  &  0.07 &  20.87 &   0.07 &  1.50 \\
 457 & 211.1  & 316.6 &  23.08  &  0.11 &  21.15 &   0.04 &  1.92 \\
 461 &  73.4  & 319.8 &  23.33  &  0.12 &  22.26 &   0.10 &  1.07 \\
 464 & 225.1  & 321.8 &  22.37  &  0.07 &  20.68 &   0.06 &  1.69 \\
 465 & 231.2  & 323.7 &  23.32  &  0.10 &  22.20 &   0.13 &  1.12 \\
 467 &  36.9  & 324.4 &  22.31  &  0.12 &  21.58 &   0.10 &  0.73 \\
 471 &  48.5  & 325.9 &  22.43  &  0.07 &  21.35 &   0.07 &  1.08 \\
 478 & 144.2  & 331.5 &  23.59  &  0.14 &  22.12 &   0.10 &  1.48 \\
 485 &  45.8  & 336.9 &  23.22  &  0.13 &  22.28 &   0.11 &  0.94 \\
 488 & 142.9  & 340.0 &  24.18  &  0.17 &  23.05 &   0.21 &  1.12 \\
 499 & 183.8  & 349.1 &  22.58  &  0.07 &  21.77 &   0.07 &  0.81 \\
\hline							    
\end{tabular}						    
\end{table}

Figure 5 shows the color-magnitude diagram of the UGC 685. The left
part shows all 209 detected points sources and indicates the 50\%
completeness limit while the right panel presents only those 173 stars
where the photometric errors are less than 0.2 in B and R. The open
symbols indicate the 136 sources which belong to the field around UGC
685 while the filled dots indicate the 73 sources in UGC 685. 56 of
the UGC 685 objects have photometric errors less than 0.2 in both bands.

A comparison with Geneva evolutionary tracks (fig. 6) as well as with
the Padua isochrones (Meynet et al, 1994; Bertelli et al, 1994)
indicates that several blue supergiants and fewer yellow supergiants
have been resolved in the mass range 15 to 60 \Msolar\
(ZAMS). Given the higher field contamination, is difficult to detect a
red supergiant population. Again, one should remember that some of
these objects, especially of the blue ones, might be unresolved
clusters. Having only one color in most cases, clusters (or blends)
can not be distinguished. Naturally, clusters of blue stars can be
brighter than individual stars. Contrary, HII regions have
been identified unambiguously with the H${\alpha}$ image.

\begin{figure}
\centerline{\psfig{figure=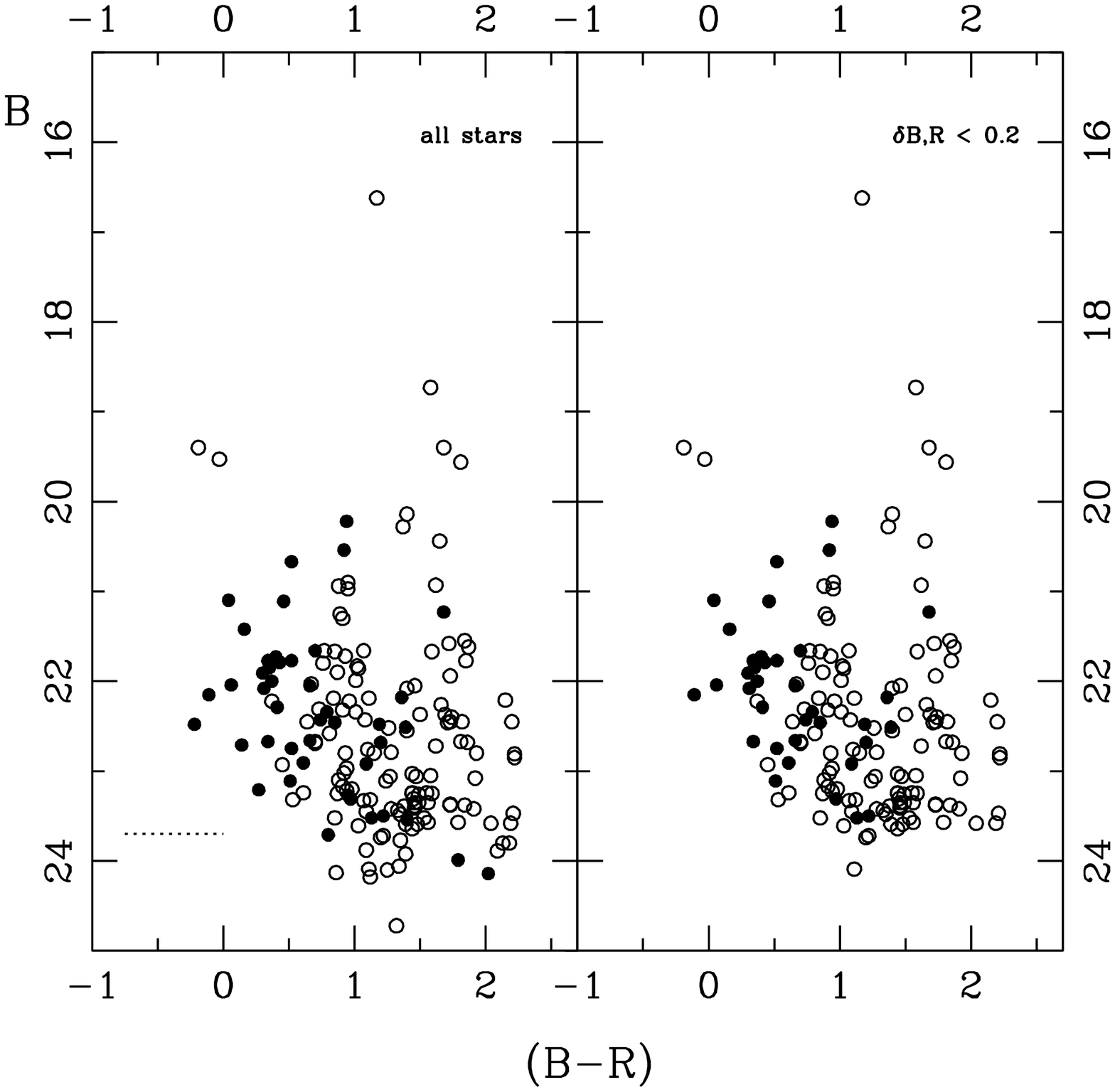,width=8.5cm,angle=0}} 
\caption[]{Left: B-R, B color-magnitude diagram of the 209 stars in
the field of UGC 685. The 136 stars outside the galaxy are marked as
circles, those 73 of UGC 685 (or projected on UGC 685) are marked by
filled dots. The dotted line indicates the 50\% completeness limit
(see text).  Right: The panel shows only those 173 sources which have
photometric errors in B and R less than 0.2. 56 of these
sources belong to UGC 685.  UGC 685 clearly shows a population of blue
and yellow supergiants pointing to recent star formation
activity. Some red supergiants seem be present too, but are harder to
disentangle from the field confusion.}
\end{figure}

\begin{figure}
\centerline{\psfig{figure=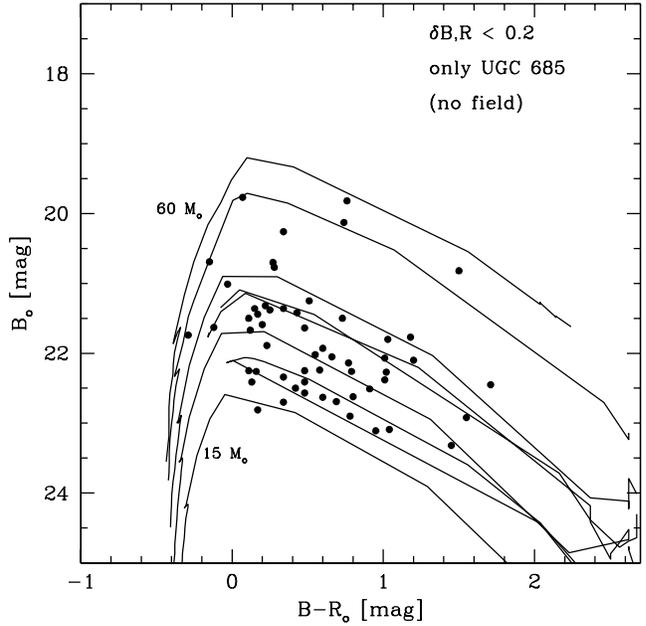,width=8.5cm,angle=0}} 
\caption[]{B-R, B color-magnitude diagram of 56 stars of UGC
685. Stars of the surrounding field and those with errors larger than
0.2 are excluded. The observations are corrected for the foreground
reddening of the Milky Way. Geneva tracks for 15, 20, 25, 40, and 60
\Msolar\ stars of a metallicity of Z = 0.001 are shown. The tracks were
taken from the http version of Schaller et al (1992) and converted
from mass, temperature, and luminosity to the observers frame with
Kurucz (1992) stellar atmospheres (where the tables with log g, T, and
colors are taken again from the web) and finally shifted to a distance
of 5.5 Mpc.}
\end{figure}

\subsection{The distance of UGC 685}

The brightest blue supergiants (BBSG's) have a long tradition
as distance indicators (K\&T, R\&RR). It is usual to determine the
mean of the three brightest blue supergiants, $B_3$, and apply a
calibration as provided by K\&T and R\&RR to derive the distance
modulus of the considered galaxy. The detailed analysis of R\&RR clearly
shows that one is limited to an accuracy of 0.8 mag. at best. Thus,
$B_3$ can give only a first estimate of the true distance. This still
excludes uncertainties which are introduced by the selection of the
candidates as interlopers from the fore- and background as well as
unresolved blends or clusters. Furthermore, the luminosity of the BBSG's
depends on the luminosity of the host galaxy. This dependance has been
explained through numerical simulations as a statistical effect
(Greggio, 1986, and references therein).  The simulation further showed that
deviations from the mean relation between $B_3$ and $M_{gal}$ up to
$\pm$ 1 mag. can be expected in the B band for galaxies which are as
faint as UGC 685. This explains the rather low accuracy of the method
as derived from empirical data by R\&RR. The errors given in the
following are based on the derivation by R\&RR.

I identified the four brightest candidates for supergiants (No. 146,
194, 217, and 250 in table 3) where I used B-R $<$ 0.4 as color
selection criterion for blue supergiants, approximating the B-V color
selection of K\&T.  The three brightest candidates yield a mean
magnitude of $B_3$ = 20.9 $\pm$ 0.6. To get an estimate of the uncertainty
introduced by clusters or blends, I also calculated $B_3$ from the second to
fourth brightest blue objects and got 21.4 $\pm$ 0.3. The values
corrected for the Milky Way foreground extinction are 20.6 and 21.1,
respectively.  Following R\&RR, their relation and fig. (10f), and
using the above values for $B_3$ and total magnitude after reddening
correction, I obtain an absolute magnitude for the brightest blue
supergiants of about -7.9 and a distance modules (distance) of 28.7
(5.5 Mpc).  According to R\&RR, the minimum error of a distance
modulus derived from their relation (10f) is 0.88. To this value,
I added quadratically the error of $B_3$ and the - almost negligible -
error in total magnitude. This yields a total error of 1.0 in
distance modulus or about 35\% in distance. The derived distance value was
used to convert observed to absolute values as given in table 2 and
to adjust the tracks (fig. 6). Following the calibration of K\&T
yields almost identical values.

I was able to identify four individual sources in the J frame which
could be cross-correlated with the objects from the B and R frames
(table 4). The colors were corrected for the Milky Way foreground
reddening. Assuming that these four objects are supergiants and that
UGC 685 itself does not contribute internal reddening, one can
translate the position in the B-R$_c$, B-J color-color diagram to a
spectral type using the table of Johnson (1966). The Johnson table
was transfer into Cousins R with Bessels (1987) relation. Using the
blue absolute magnitudes of supergiants for these spectral types as
listed by Schmidt-Kaler (1982), one gets distance moduli as also given
in table 4. Three objects combine to m-M = 28.8 (5.8 Mpc), including
the fourth one yields 29.1 (6.7 Mpc). I take this as a consistency
check for the B$_3$ method. The deviating result for the fourth object
shows the limitation of this approach and may be caused by a
blend or one of the above assumptions may not be valid in this case.

\setcounter{table}{3}
\begin{table}
\caption{Individual objects identified in B, R, and J
within UGC 685. Listed are the numbers from table 3, magnitudes and
colors, the spectral types as estimated from the color-color values, and
the individual dereddened distance moduli assuming that the objects
are supergiants in UGC 685. See text for details. }
\begin{tabular}{l|c|c|c|c|c}
\hline
star number & B & B-R & B-J & Spectral & m-M$_{0}$ \\
from table 3&   &     &     & type     &        \\ 
\hline
155 & 20.54 & 0.93 & 0.66 & F1I: & 28.78\\
176 & 22.18 & 1.36 & 1.47 & F9I  & 30.12\\
207 & 20.67 & 0.52 & 0.43 & A5I  & 28.91\\
216 & 21.33 & 1.68 & 2.18 & G5I  & 28.71\\
\end{tabular}
\end{table}

The two distance estimates presented here are in rather good agreement
with an estimate based on the observed HI velocity and a Virgo infall
model which yields about 6.0 Mpc (Schmidt \& Boller, 1992a). Thus, UGC
685 really belongs to the 10 Mpc sample and is quite isolated in
space.

\subsection{Star formation history}

According to the results described so far, UGC 685 is in the right
distance for an HST multicolor study which would provide the data for
a detailed reconstruction of the star formation history 
(Schulte-Ladbeck et al. 1998, see also the review by Grebel,
1997). Even though the ground-based data presented in this paper have a much
more limited resolution, they already indicate that the recent star
formation rate (SFR) is rather low and that it varied only
slightly during the past.

According to the results in chapter 3.2 and 3.3, the observed total
H${\alpha}$ flux corresponds to a recent (within the last $10^7$ yr)
SFR of 0.003 \Msolar yr$^{-1}$. As shown, this
activity is spatially highly concentrated. The resolved supergiants
and the underlying light distribution are related to slightly older
star formation activities.  The resolved supergiants are almost all inside
26~arc~sec (major axis) which corresponds to 0.70~kpc or 2 scale
lengths of the overall light distribution. This indicates that averaged
over a longer time interval of $\sim 10^8$ yr, the SF took place
all over the central part of UGC 685. This is supported by the absence
of significant color gradients in B-R and B-J and the very limited
deviation from the average in the color map.

A further estimator of the SFR of an actively star forming galaxy is
its total blue luminosity. As described by Gallagher et al. (1984),
this estimator $\alpha_L$ is strictly valid only for a constant SFR
and depends on the assumed IMF and its upper and lower mass
cut-offs. It also slightly depends on the chemical composition of the
system (see Gallagher et al., 1984, for complete references). The blue
light is dominated by stars of $2.5~\Msolar > {\rm M}_\ast > 1$~\Msolar\ 
($ 4\cdot10^8 yr > {\rm age} > 6\cdot10^9$ yr). 
Applying the Gallagher et al. formula (7) yields the
same SFR as derived from the H$\alpha$ flux. Thus, I have little
evidence for strong variations of the SFR with time up to $\sim 10^9$ yr.

To strengthen this point, I used the total optical colors B-R
and B-I (table 2). I compared the observed, dereddened colors with
Bruzual \& Charlot (1994) models with exponential declining star
formation rate of e-folding time $\tau$. Models with $\tau \le 5$ Gyr
do not fit the colors at all and are far too red. A $\tau = 10$ Gyr model
can reproduce the colors rather well with deviations of less than 
0.1\footnote{A formation time of $z_f = 4$, h = 0.6, and $q_0 =
0.1$ are assumed; B-J is not used because of some uncertainties of the
NIR model colors.}. Models with an exponential decay and a recent, even
small, burst are in conflict with the data. Therefore, also the colors
support the scenario of a simple star formation history.

In other words, the isolated dwarf irregular galaxy UGC 685 seems to
have an approximately constant and low SFR, roughly stable for the past 
$10^9$ yr. The SFR is also quite low when compared to other galaxies
of the same morphological type -- actually, UGC 685 ranges at the lower
end of measured SFR values of irregular galaxies. As I
applied the formulae of Gallagher et al (1984), Hunter \& Gallagher
(1985, 1986), it is straightforward to compare the obtained SFR's of UGC
685 to the values for their sample galaxies. For dwarf irregular
galaxies, they found SFR values between 0.5 and 0.0002 \Msolar
yr$^{-1}$ (giant irregulars: 0.3 to 1.1). The SFR correlate with the
absolute blue magnitude. Galaxies which have the similar M$_B \pm 0.5$
as UGC 685 show similar SFR values.

As indicated by the amount of detected HI gas, this galaxy can
continue with star formation at its today rate for quite a long
time. To do so, the reservoir of far outlying HI gas has to be moved
into those central regions where conditions obviously support star
formation. As no external trigger seems to be present, this might be a
very slow process, depending on the dynamical evolution of the
extended HI disk. It might even be possible that the stellar body of
todays UGC 685 will already be totally dimmed to an object resembling
a dwarf spheroidal before this outlying gas is able to move in. In
this context, it is interesting to note that Carignan et al. (1998)
found some outlying HI gas in the Sculptor dwarf spheroidal galaxy.

\section{Conclusions}

Optical broad and narrow band as well as NIR imaging data were
presented and used to derive the structural parameters of the isolated
dwarf galaxy UGC 685 which show very little signs of
irregularities. UGC 685 is one of those dwarfs where the HI gas is
reaching to far greater distances than the stars, and might be,
like DDO 154, a good case for dark matter halo studies. In a separate
paper, I will show that the stellar and ionized gas kinematics fit
quite well to the HI rotation curve established by Hoffman et al
(1996).  Total fluxes and colors, the color distribution, and the
especially H$\alpha$ flux and morphology point to a low star formation
rate. Even for dwarf irregular galaxies, it is at the lower boundary of
the observed values (Hunter\& Gallagher, 1985, 1986). The overall star
formation history seems to be calm at a level of 0.003 \Msolar
yr$^{-1}$ over the last $\sim 10^9$ yr, but this particular irregular
dwarf was nevertheless able to establish a surface brightness at the
upper boundary of dwarfs of this size (compare to Binggeli, 1993, his
Fig. 4, or Hopp, 1994, Fig. 1). Over short intervals ($\sim 10^7$ yr),
the star formation activity appears strongly localized.  Data with
much better resolution are needed for a more detailed study and
reconstruction.

A new distance estimate was established based on the resolved
brightest supergiants. This new estimates confirms that UGC 685
belongs to the 10 Mpc sample of the very nearby universe and is indeed
a very isolated galaxy.

\acknowledgements
I would like to thank the Calar Alto staff for his kind support during the
observations. Drs. Ralf Napiwotzki and Sabine Moehler helped with the
electronic version of the Kurucz tables. I acknowledge many useful
discussions with Drs. Bender, Greggio, Rosa, and Schulte-Ladbeck.  
Niv Drory did the Bruzual \& Charlot model calculations. 
I was supported by the DFG (hopp/1801-1) and by the SFB 375.

{}

\begin{thebibliography}{}

\bibitem{}
Babul,A., Ferguson,H., 1996, ApJ 458, 100

\bibitem{}
Bender,R., M\"ollenhoff,C., 1987, A\&A 177, 71

\bibitem{}
Bessel, M.S., 1987, PASP 99, 642

\bibitem{}
Bertelli,G., Bressan,A., Chiosi,C., Fagotto,F., Nasi,E., 1994, A\&AS 106,275

\bibitem{}
Binggeli, B., 1993, in: 'Panchromatic View of Galaxies', G. Hensler,
Ch. thies, J.S. Gallagher (eds.), Edition Frontieres, p. 173

\bibitem{}
Bomans,D., Chu,Y.H., Hopp, U., 1996, AJ 113, 1678

\bibitem{} 
Bruzual, A.G., Charlot, S., 1993, ApJ 405, 538

\bibitem{} 
Burstein D., Heiles C., 1982, AJ 87, 1165

\bibitem{} 
Carignan, C., Beaulieu, S., Cote, S., Demers, S., Mateo, M., 1998, AJ
in press (astro-ph/9807222)

\bibitem{} 
Christian C.A., Adams M., Barnes J.V., Butcher H., Mould
J.R., Siegel M., 1985, PASP 97, 363  

\bibitem{}
Elias,J.H., Frogel,J.A., Matthew,K., Neugebauer,G., 1982, AJ 87, 1029

\bibitem{}
Ferguson,H., Babul,A., 1998, MNRAS, preprint, astro-ph/9801057

\bibitem{} 
Gallart,C, Aparicio,A., Vilchez,J.M., 1996, AJ 112, 1928

\bibitem{} 
Gallagher, J. S., Hunter, D. A., Tutukov, A.V., 1984, ApJ 284, 544

\bibitem{}
Grebel, E.K., 1997, Rev. in Modern Astron. 10, 29

\bibitem{} 
Greggio,L., 1986 A\&A 160, 111

\bibitem{} 
Greggio,L., 1994, ESO Conf. Workshop Proc. 51, 72

\bibitem{}
Hoffman,G.L., Salpeter,E.E., Farhat,B., Roos,T., Williams,H.,
Helou,G., 1996, ApJS 105, 269

\bibitem{}
Hopp,U., 1994, ESO Conference \& Workshop Proc. 49, 37

\bibitem{}
Hopp,U., Schulte-Ladbeck,R.E., 1991, A\&A 248, 1

\bibitem{}
Hunter,D., 1997, PASP 109, 937

\bibitem{}
Hunter, D. A., Gallagher, J. S., III, 1985, ApJS 58, 533

\bibitem{}
Hunter, D. A., Gallagher, J. S., III, 1986, PASP 98, 5

\bibitem{}
Hunter, D. A., Hawley, W. N., Gallagher, J. S., III, 1994, AJ 106, 1797

\bibitem{}
Hunter,D., Elmegren,B.G.., Backer,A.L., 1998, ApJ 493, 595

\bibitem{}
Johnson, H.L., 1966, ARA\&A 4, 193

\bibitem{} 
Karachentsev,I.D., Tikhonov, N.A., 1993, ESO Conf. Workshop Proc. 49,
109 (K\&T)

\bibitem{}
Kran-Korteweg,R., 1986, A\&AS 66, 255

\bibitem{}
Kran-Korteweg,R., Tammann,G.A., 1979, AN 300, 181

\bibitem{}
Kurucz, R.L., 1992, in: The Stellar Populations of Galaxies,
eds. B. Barbuy \& A. Renzini, Dordrecht, Kluwer, p. 255

\bibitem{}
Lu,N.Y., Hoffman,G.L., Groff,T., Roos.T., Lamphier,C., 1993, ApJS 88, 383

\bibitem{}
Meynet,G., Maeder,A., Schaller,G., Schaerer,D., Charbonnel,C., 1994
A\&AS 103, 97

\bibitem[1994]{RRR94}
Rozanski,R., Rowen-Robinson,M., 1994, MNRAS 271, 530 (R\&RR)

\bibitem{} 
Schaller, G., Schaerer, D., Meynet, G., Maeder, A., 1992 A\&AS 96, 269

\bibitem{} 
Schmidt,K.H., Boller, Th., 1992a, AN 313, 190

\bibitem{} 
Schmidt,K.H., Boller, Th., 1992b, AN 313, 330

\bibitem{}
Schmidt-Kaler,Th., 1982, in: {\it Landolt-B\"ornstein}, Numerical Data
and Functional Relationships in Science and Technology, N.S. VI,
2b, ed. K.H. Hellwege, Springer-Verlag, Berlin

\bibitem{}
Schulte-Ladbeck,R.E., Crone,M.M., Hopp,U., 1998, ApJ 493, L23

\bibitem{}
Taylor, C.L., Brinks, E., Grashuis, R.M., Skillman, E.D., 1996 ApJS
102, 189

\bibitem{}
van Zee, L., Skillman, E. D., Salzer, J. J., 1998, AJ in press
(astro-ph/9806246) 

\end{thebibliography}
\end{document}